\documentclass[prl,twocolumn,showpacs,superscriptaddress]{revtex4}
\usepackage{amsmath,amsfonts,graphicx}

\begin{document}

\title{Oscillations in the expression of a self-repressed gene\\
   induced by a slow transcriptional dynamics.}

\author{Pierre-Emmanuel Morant}
\author{Quentin Thommen}
\affiliation{Universit\'e des Sciences et Technologies de Lille,
  PhLAM,  F-59655   Villeneuve d'Ascq, France.}
\affiliation{CNRS, UMR 8523, FR 2416, F-59655 Villeneuve d'Ascq, France.}
\author{Fran\c{c}ois Lemaire}
\affiliation{Universit\'e des Sciences et Technologies de Lille,
  LIFL, 59655   Villeneuve d'Ascq, France.}
\affiliation{CNRS, UMR 8022, F-59655 Villeneuve d'Ascq, France.}
\author{Constant Vandermo\"ere}
\affiliation{Universit\'e des Sciences et Technologies de Lille,
  PhLAM,  F-59655   Villeneuve d'Ascq, France.}
\affiliation{CNRS, UMR 8523, FR 2416, F-59655 Villeneuve d'Ascq, France.}
\author{Benjamin Parent}
\affiliation{Universit\'e des Sciences et Technologies de Lille,
  UGSF, F-59655 Villeneuve d'Ascq, France.}
\affiliation{CNRS, UMR 8576, F-59655 Villeneuve d'Ascq, France.}
\affiliation{Interdisciplinary Research Institute,  USR CNRS 3078,
  Universit\'e des Sciences et Technologies de Lille, F-59655
  Villeneuve d'Ascq, France.}
\author{Marc Lefranc}
\affiliation{Universit\'e des Sciences et Technologies de Lille,
  PhLAM,  F-59655   Villeneuve d'Ascq, France.}
\affiliation{CNRS, UMR 8523, FR 2416, F-59655 Villeneuve d'Ascq, France.}
\affiliation{Interdisciplinary Research Institute,  USR CNRS 3078,
  Universit\'e des Sciences et Technologies de Lille, F-59655
  Villeneuve d'Ascq, France.}

\date{\today}

\begin{abstract}
  We revisit the dynamics of a gene repressed by its own protein in
  the case where the transcription rate does not adapt instantaneously
  to protein concentration but is a dynamical variable. We derive
  analytical criteria for the appearance of sustained oscillations and
  find that they require degradation mechanisms much less nonlinear
  than for infinitely fast regulation. Deterministic predictions are
  confirmed by stochastic simulations of this minimal genetic
  oscillator.
\end{abstract}


\maketitle

Networks of genes interacting via regulatory proteins modulating their
activities are highly nonlinear systems which display a variety of
dynamical
behaviors~\cite{Goldbeter96book,Fall02:_comput_cell_biolog,tiana07:_oscil}.
Their modeling has generally assumed that gene activation is fast
compared to other processes so that transcription rate reacts
instantaneously to protein concentration. However, transcription is a
complex process~\cite{xavier07:_in_rna_ii}. In the last years, it has
been increasingly recognized that gene activity fluctuations can be
slow and that this can affect the behavior of gene regulatory
networks. In particular, slow transcriptional bursting and
transcriptional memory have been observed
experimentally~\cite{Golding05:_real_gene_indiv,%
  chubb06:_trans_pulsin,raj06:_stoch}. 
Theoretically, it has been shown that slow activation dynamics can
lead to bursts in
expression~\cite{hornos05:_self,zon06:_diffus_trans}, induce
bistability~\cite{lipshtat06:_genet_toggl} or modify the flipping rate
of a genetic switch~\cite{walczak05:_absol,morelli08:_react_coord}.

In this Letter, we show that slow promoter dynamics can also lead to
oscillations by investigating how transcriptional dynamics modifies
the behavior of a single gene repressed by its own
protein~\cite{Goodwin65a,Griffith68a,Bliss82a,Goldbeter95a,%
  Leloup99a,Jensen2003a,Monk2003a,Lewis2003a}. This old problem of
theoretical biology has been recently revived by the study of the
\emph{Hes1} gene involved in the somite clock~\cite{Hirata02a}. The
usual view is that oscillations appear in this genetic circuit only
when additional steps are inserted in the feedback
loop~\cite{Griffith68a,tiana07:_oscil}. In the Goodwin and Bliss
oscillators~\cite{Goodwin65a,Griffith68a,Bliss82a}, the gene protein
catalyzes synthesis of the actual repressor. In early circadian
models, transport of the repressor into the
nucleus~\cite{Goldbeter95a,Leloup99a} is a key
oscillatory ingredient. In fact, the mere introduction of a time delay
in the one-gene circuit model (accounting for protein transport or
more generally a cascade of intermediate steps~\cite{Morelli07a}) can
destabilize it~\cite{Jensen2003a,Monk2003a,Lewis2003a,tiana07:_oscil}.
As will be of particular interest here, oscillations may also be
induced by strongly nonlinear degradation
mechanisms~\cite{Tyson99:_per_tim}.

For simplicity, we study the case of an elementary kinetic equation
describing regulation through protein-DNA
binding~\cite{Francois05:_core}, which we however prefer to view as a
minimal description of transcriptional memory in more complex
mechanisms. We derive an analytical expression of the oscillation
threshold, and show that when the gene response time is appropriately
tuned, the one-gene circuit can be destabilized (and oscillations
induced) by degradation mechanisms much less nonlinear than for
infinitely fast regulation. This result provides new insights into the
interplay of nonlinearity and time delay. Stochastic simulations
confirm that the main results of our analysis carry over to low copy
number situations.

\begin{figure}[htbp]
  \centering
  \includegraphics[width=3in]{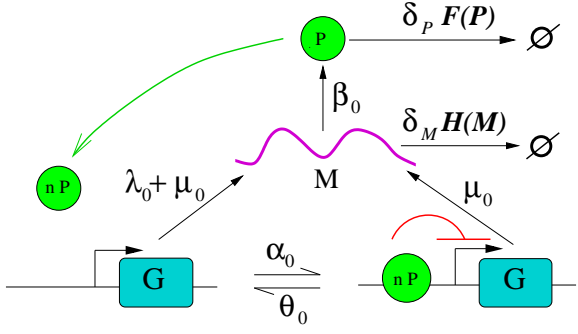}
  \caption{Reaction diagram of the self-regulated gene circuit.}
  \label{fig:model}
\end{figure}

Our study is based on the following three-variable model describing
the genetic circuit represented in Fig.~\ref{fig:model}:
\begin{subequations}
\begin{eqnarray}
  \dot{G} &=& \theta_0 (1-G) - \alpha_0 C(P)G\label{eq:modelgene}\\
  \dot{P} &=& n\dot{G}+\beta_0 M-\delta_P
  F(P)\label{eq:modelprot}\\ 
  \dot{M} &=& \mu_0+\lambda_0 G - \delta_M H(M)\label{eq:modelarn}
\end{eqnarray}
  \label{eq:model}
\end{subequations}
where $G$, $P$ and $M$ represent gene activity, protein and RNA copy
numbers. Eq.~\eqref{eq:modelgene} formally describes the kinetics of
protein-DNA binding at rate $\alpha_0$ and unbinding at rate
$\theta_0$~\cite{Francois05:_core}. Possible cooperativity effects are
taken into account via the function $C(P)$ and the number $n$ of
proteins binding DNA. Single-protein regulation corresponds to
$C(P)=P$, $n=1$. More generally, Eq~\eqref{eq:modelgene} is a minimal
model for the dynamics of an effective gene activity $G$ slowly
relaxing towards an equilibrium value given by the gene regulation
function $G=1/[1+C(P)/C(P_0)]$, with $P_{0}$ the half-expression
threshold. Such a model can be obtained as the leading approximation
of a mechanistic model of transcription including all processes
concurring to gene expression (e.g., protein-DNA interaction,
formation of open complex, polymerase recruitment, chromatin
remodeling,...) when there is a dominant limiting step.
Eq.~(\ref{eq:modelgene}) causes gene activity $G$ to lag fast changes
in protein level, and plays a dynamical role similar to explicit time
delays~\cite{Jensen2003a,Monk2003a,Lewis2003a} or to transport
equations~\cite{Leloup99a}. In Eq.~\eqref{eq:modelprot}, the three
terms correspond to binding/unbinding, translation and degradation.
Eq.~\eqref{eq:modelarn} describes transcription at rate
$\mu_0+\lambda_0G$ and RNA degradation. In order to understand how
oscillations can be induced by tuning protein and RNA degradation, we
derive the oscillation criterion for arbitrary degradation functions
$F(P)$ and $H(M)$ with unit derivative at zero, $\delta_P$
and $\delta_M$ being the low-copy-number degradation rates.

By
renormalizing time, variables, parameters, cooperativity and
degradation functions according to:
\begin{subequations}
  \begin{eqnarray}
    t = \frac{t'}{\delta_M},\;
    G = g,\; P = pP_0,\; M=m M_0,\; \\
    P_0=C^{-1}\left(\frac{\theta_0}{\alpha_0}\right),\; M_0=
    \frac{\delta_PP_0}{\beta_0},\;
    \theta=\frac{\theta_0}{\delta_M},\;\\
    \alpha=\frac{\theta_0}{P_0\delta_M},\;
    \delta=\frac{\delta_P}{\delta_M},\;
    \lambda=\frac{\lambda_0}{M_0\delta_M},\;
    \mu=\frac{\mu_0}{M_0\delta_M}
    \\
     c(p) = \frac{C(P)}{C(P_0)},\,f(p)=\frac{F(P)}{P_0},\;
    h(m)=\frac{H(M)}{M_0},
  \end{eqnarray}
\end{subequations}
Eqs.~\eqref{eq:model} can be rewritten in dimensionless form
\begin{subequations}
\begin{eqnarray}
  g' &=& \theta \left[1-g(1+c(p))\right]\label{eq:redmodelgene}\\
  p' &=& n\alpha \left[1-g(1+c(p))\right] + \delta[m- f(p)]\\
  m' &=& \mu+\lambda g - h(m)
\end{eqnarray}
\label{eq:redmodel}
\end{subequations}
where $x'=dx/dt'$. When $f$ and $g$ are monotonous and
$h[f(\infty)]>\mu$, model~\eqref{eq:redmodel} has a single steady
state $(g_*,p_*,m_*)$ satisfying the fixed point equations:
\begin{equation}
  \label{eq:fixedpoint}
  g_*=\frac{1}{1+c(p_*)},m_* = f(p_*), g_*=\frac{h(m_*)-\mu}{\lambda}.
\end{equation}

Note that the steady state depends only on parameters $\lambda$ and
$\mu$ as well as on functions $c$, $f$ and $h$, whereas
parameters $\theta$, $\alpha$, $\delta$ control time scales. The
behavior of the degradation and cooperativity functions in the
neighborhood of the steady state is described by the slopes
\begin{displaymath}
s= \left.\frac{\mathrm{d}f(p)}{\mathrm{d}p}\right|_{p=p_*},\,
u=\left.\frac{\mathrm{d}h(m)}{\mathrm{d}m}\right|_{m=m_*},\,
v=\left.\frac{\mathrm{d}c(p)}{\mathrm{d}p}\right|_{p=p_*}
\end{displaymath}
In the case of linear degradation [$f(p)=p$, $h(m)=m$], we have
$u=s=1$. Small or even negative values of the slopes $s$ and $u$
generally denote strongly nonlinear degradation
mechanisms~\cite{Fall02:_comput_cell_biolog,Tyson99:_per_tim},
including saturation.

To assess whether Eqs.~\eqref{eq:redmodel} can display oscillations,
we have searched for parameter values where the fixed point specified
by~\eqref{eq:fixedpoint} loses stability to a periodic solution via a
Hopf bifurcation (i.e., a pair of conjugate eigenvalues of the
linearized problem cross the imaginary axis). For simplicity, we
assume perfect repression ($\mu=0$) and a large threshold $P_0$
($\alpha\sim 0$). Under this approximation, the Routh-H\"urwitz
stability criterion~\cite{gradshteyn00} indicates that a Hopf
bifurcation occurs when the quantity
\begin{equation}
  \label{eq:routh}
  \mathcal{H}= \sigma+
   \left ( -\delta \lambda v g_*^2+ \sigma^2 \right )\tau
  +  \gamma  \sigma \tau^2
\end{equation}
crosses zero to become negative, where $\tau=g_*/\theta$ is the gene
response time, and the sum $\sigma = \delta s+u$ and product $\gamma =
\delta s u$ are symmetric functions of degradation rates $\delta s$
and $u$. At bifurcation, the oscillation period is
$\tau_\text{osc} = 2\pi\sqrt{\tau/(\sigma+\gamma \tau)}$ where
$\sigma$ and $\gamma$ satisfy $\mathcal{H}=0$. 
Cooperativity essentially changes feedback strength from
$\delta\lambda$ to $\delta\lambda v$, thus we assume for simplicity
single-protein regulation [$c(p)=p$, $v=1$] thereafter.

Eq.~(\ref{eq:routh})
shows that a strong feedback destabilizes the system while high
degradation rates (large $\sigma$ and $\gamma$) tend to stabilize it.
In the single protein case, $\mathcal{H}>0$ when both protein and RNA
are linearly degraded and no oscillations occur. Conversely, when
protein and RNA degradations are completely saturated ($s = u=
\sigma=0$), $\mathcal{H} = -\delta\lambda g_{*}^2\tau < 0$, indicating
that oscillations then appear systematically. The behavior in
intermediate cases depends on the value of the response time $\tau$.

In the classical case $\tau=0$, $\mathcal{H}= \sigma$ and oscillations
appear only for $\sigma<0$. It is indeed known that negative effective
degradation rates can lead to
oscillations~\cite{Fall02:_comput_cell_biolog,Tyson99:_per_tim}. We
thus restrict ourselves to showing that at finite $\tau$, oscillations
can occur for $u,s>0$. More precisely, we want to understand how
oscillations arise away from the saturated cases $u=0$ or $s=0$. To
this end, we use the geometric slope average $\nu = \sqrt{us} =
\sqrt{\gamma/\delta}$ as an index ($\nu=1$ in the linear case), seeking
to determine the maximum value of $\nu$ at which oscillations can be
observed, and for which values of $\tau$ this extremum is achieved.

The quantities $\sigma$ and $\gamma$ play complementary roles. The total
degradation rate $\sigma$ controls instability onset for small to
moderate $\tau$. Moreover, Eq. (\ref{eq:routh}) indicates that when
$\sigma,\gamma>0$, a necessary condition for oscillations is
\begin{equation}
  \label{eq:necessary}
  \sigma < \sigma_c = g_* \sqrt{\delta\lambda} =
  \frac{g_*\sqrt{2}}{t_\text{sw}} 
\end{equation}
where $t_\text{sw}$ is the time during which a fully active gene
synthesizes the amount of protein corresponding to half-repression
threshold. The degradation rate product $\gamma$ is
relevant only for large $\tau$, blocking oscillations if it is too
large. In particular, $\gamma=0$ guarantees the onset of oscillations for
sufficiently large $\tau$ whenever~(\ref{eq:necessary}) holds. For
$u,s>0$ and a given value of $\sigma$, $\gamma$ can take any value
between $\sigma^2/4$ and $0$ depending on whether the two degradation
rates are equal or completely unbalanced, one being equal to zero and
the other to $\sigma$. We use below $\epsilon=2\sqrt{\gamma}/\sigma \in
[0,1]$ as a balance indicator.

\begin{figure}[htbp]
  \centering
  \includegraphics[width=3.4in]{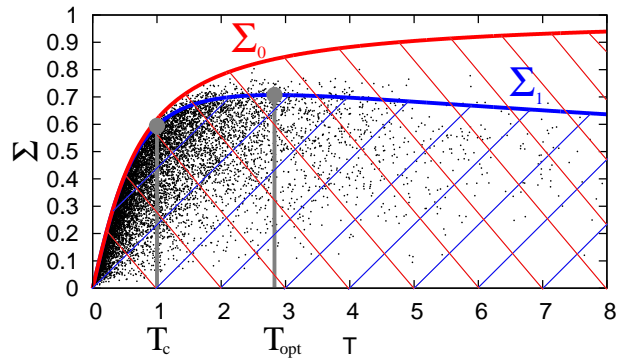}
  \caption{Bifurcation diagram of Eqs.~(\ref{eq:redmodel}) in the
    $(\Sigma,T)$ plane according to (\ref{eq:routhred}). A system with
    balance index $\epsilon$ oscillates for
    $\Sigma<\Sigma_\epsilon(T)$ (hatched areas). Black dots indicate
    oscillating parameter sets of ~(\ref{eq:redmodel}), with
    $\theta,\delta\in[10^{-1},10]$, $\theta/\alpha\in[10,1000]$,
    $\lambda\in[1,10^3]$, $\lambda/\mu\in[10,10^3]$, assuming the
    degradation mechanisms~(\ref{eq:degradex}) with $a\in[10^{-6},1]$,
    $\kappa,\chi\in[1,100]$.
    $T_c=1$ is the time scale at which transcriptional dynamics
    cannot be neglected, $T_{opt}=2\sqrt{2}$ is the location of the
    maximum of $\Sigma_1(T)$.}
  \label{fig:diagbif}
\end{figure}

Remarkably, we note that under the rescaling
\begin{equation}
  \label{eq:scaling}
 \sigma = \sigma_c \Sigma,\quad
 \gamma = \sigma_c^2\left(\frac{\epsilon\Sigma}{2}\right)^2,
 \quad \tau = \frac{T}{\sigma_c},
\end{equation}
the oscillation condition can be
rewritten without explicit parameter dependence:
\begin{equation}\label{eq:routhred}
  \mathcal{H}_\epsilon(\Sigma,T) = \Sigma\times \left[ \frac{\epsilon^2\Sigma^2}{4} T^2+\left(\Sigma-\frac{1}{\Sigma}\right)T+1\right] < 0
\end{equation}
and defines a series of curves $\Sigma_\epsilon(T)$ such that given a
balance index $\epsilon$ and a response time $T$, oscillations are
found for $\Sigma\le\Sigma_\epsilon(T)$. Fig.~\ref{fig:diagbif} shows
the limit curves $\Sigma_1(T)$ and $\Sigma_0(T)$ which are important
to understand the bifurcation diagram: regardless of the value of
$\epsilon$, the circuit always (resp., never) oscillates when
$\Sigma<\Sigma_1(T)$ [resp., $\Sigma>\Sigma_0(T)$]. To support our
analysis, we have searched the parameter space of
Eqs.~(\ref{eq:redmodel}) for oscillatory behavior for $\alpha,\mu\neq
0$, assuming for definiteness allosteric protein degradation and
Michaelis-Menten RNA degradation:
\begin{equation}\label{eq:degradex}
  f(p) = \frac{p\times (a+p/\kappa)}{a+2a(p/\kappa)+(p/\kappa)^2},\;
  h(m) = \frac{\chi m} {\chi + m}
\end{equation}
Points in the $(\Sigma,T)$ plane associated with oscillating parameter
sets are shown as black dots in Fig.~\ref{fig:diagbif}. Agreement is
excellent: all dots are below the $\Sigma_0(T)$ curve and the few
significantly above $\Sigma_1(T)$ have one small degradation rate. We
are thus confident that our analysis allows us to understand the
behavior of Eqs.~(\ref{eq:redmodel}).

Two regions can be distinguished in Fig.~\ref{fig:diagbif}. For $T<1$,
the instability threshold $\Sigma_\epsilon(T)$ is practically
independent of $\epsilon$ and increases rapidly with $T$. For small
$T$, $\Sigma_\epsilon(T)\sim T$ (thus, the oscillation criterion is
$\sigma < \sigma_c^2  \tau$). In the $T>1$ region,
$\Sigma_\epsilon(T)$ reaches its maximum value
$\Sigma_m(\epsilon)=1/\sqrt{\epsilon+1}$ at
$T=T_m(\epsilon)=2\sqrt{\epsilon+1}/{\epsilon}$, and then decreases as
$T^{-1}$ for $T\to\infty$, except for $\epsilon=0$ where it
monotonously increases towards $\Sigma=1$. At fixed $\Sigma$,
oscillations are thus found in a finite range of $T$, which widens
gradually, and is eventually infinite, as $\epsilon\to0$.

In the oscillation region, the index $\nu \sim \epsilon\Sigma$
measuring distance from saturated degradation reaches its maximal
value $\nu_\text{opt}=g_*\sqrt{\lambda/8}$ for
$T=T_\text{opt}=2\sqrt{2}$ and $\epsilon=1$, at the maximum of the
$\Sigma_1(T)$ curve. Our analysis thus unveils a resonance phenomenon
in the dynamics of a self-regulated gene with dynamical transcription
rate: this circuit bifurcates most easily to periodic behavior, or
more generally is least stable, at a finite value of the gene
relaxation time given by $\tau_\text{opt} = 2\sqrt{2}\tau_c$ where
\begin{equation}
  \label{eq:tauopt}
  \tau_c = \frac{1}{g_*\sqrt{\delta\lambda}}
  =\frac{1}{\sigma_c}
     = \frac{t_\text{sw}}{g_*\sqrt{2}}
   = \delta_M\times \frac{1}{g_*}
   \sqrt{\frac{P_0}{\lambda_0\beta_0}}.
\end{equation}
The quantity $\tau_c$ gives the time scale at which dynamical behavior
departs from the fast regulation case. Because $g_*$ is determined
by~(\ref{eq:fixedpoint}), computing precise lower bounds on $\tau_c$
with (\ref{eq:tauopt}) requires specifying the degradation mechanisms.
Fixing $\lambda_0=\beta_0=10$ mn$^{-1}$, $P_0=100$, and $g_*=0.5$,
provides an estimate $t_c =\tau_c/\delta_M = 2$ mn which is not
unrealistically larger than typical gene induction times. At
resonance, the oscillation period is
$\tau_{\text{osc}}=4\pi\sqrt{2/3}\;\tau_c$ (about 20 mn in the example
above).



Since a common interpretation of the diagram of Fig.~\ref{fig:model}
is that there are only two gene states (bound or unbound)
\cite{kepler01:_stoch}, one may wonder whether our
deterministic analysis is relevant. If $g$ is viewed as a
temporal average of gene activity, our results are valid when
the response time $\tau$ is small compared to the oscillation period
so that there are many binding/unbinding events by
cycle~\cite{forger05:_stoch}. Moreover, transcription is a complex
process involving a number of distinct
steps~\cite{xavier07:_in_rna_ii}, and Eq.~(\ref{eq:modelgene}) is the
simplest way to model memory effects arising from cooperativity in the
transcription machinery~\cite{chubb06:_trans_pulsin}.

Anyhow, we now show that even when $G$ is a stochastic variable
jumping between $0$ and $1$, our main result still holds: there is a
time scale near $\tau_\text{opt}$ at which oscillations are enhanced.
To this end we have carried out stochastic simulations of the reaction
network of Fig.~\ref{fig:model} using the Gillespie
algorithm~\cite{gillespie77:_exact}, varying the response time $\tau$
at fixed $P_0$. Instead of regular oscillations, a sequence of
irregularly spaced peaks in protein concentration is observed. A
natural question is then whether protein peaks occur more regularly at
parameter values where the deterministic model oscillates, in
particular when $\tau=\tau_\text{opt}$.

\begin{figure}[tb]
  \centering
  \includegraphics[width=3.3in]{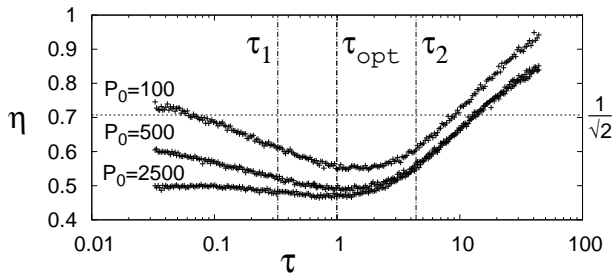}
  \caption{Coefficient of variation of interpeak time intervals $\eta$
    vs. mean residence time $\tau$ for a stochastic simulation of
    (\ref{eq:redmodel})+(\ref{eq:degradex}) rescaled to original
    variables $G$, $P$, $M$ at three values of threshold $P_0$ (from
    top to bottom, $P_0=100$, $500$, $2500$). Parameter values
    correspond to $\delta=1$, $\theta=1$, $\alpha=\theta/P_0$,
    $\lambda=21.54$, $\mu=0.085$, $\chi=95.5$, $\kappa=21.68$,
    $a=10^{-6}$. $\tau_1$ and $\tau_2$ are the boundaries of the
    deterministic oscillation domain.}
  \label{fig:stoch}
\end{figure}

We define interpeak times as the time intervals $\Delta t$ between two
crossings of $P=1.2 P_\text{avg}$ separated by at least one crossing
of $P=0.8 P_\text{avg}$, with $P_\text{avg}$ the mean protein level
(thus imposing a minimum amplitude of 40\%). Their distribution is
characterized by the coefficient of variation $\eta =
\frac{\sigma_{\Delta t}}{<\Delta t>}$. A typical variation of $\eta$
with $\tau$ in our 
system is shown in Fig.~\ref{fig:stoch}. It definitely suggests that the
deterministic analysis remains relevant in the stochastic regime,
since  there is clearly a time scale near $\tau_\text{opt}$ where
interpeak time fluctuations are minimal.

In conclusion, we have shown that a nontrivial transcriptional
dynamics can destabilize a self-regulated gene. Although it is known
that nonlinear degradation mechanisms can induce oscillations in this
system, we observe a resonance-like effect such that a much weaker
nonlinearity is required when the gene response time matches a
characteristic time. Its expression can be computed analytically,
which allows us to identify the parameter regions where this effect
cannot be neglected. Stochastic simulations confirm the relevance of
this time scale in the dynamics of the self-regulated gene. This shows
that transcriptional dynamics is a possible source of oscillatory
behavior besides other deterministic
\cite{Goldbeter96book,Fall02:_comput_cell_biolog,tiana07:_oscil,Goldbeter95a,Leloup99a,%
  Jensen2003a,Monk2003a,Lewis2003a,Tyson99:_per_tim} and
stochastic~\cite{loinger:051917,blossey:2008} effects.

A natural question is whether our conclusions remain valid when more
detailed transcriptional mechanisms or multiple sources of delay are
taken into account. It can be shown that adding an explicit delay to
Eqs.~(\ref{eq:redmodel}) only further destabilizes the circuit and
thus does not essentially interfere with the effect of a gene response
delay. Future studies should thus focus on cases where the
transcriptional dynamics is more complex and features more than one
limiting step.

\end{document}